\pdfoutput=1 
\documentclass[sn-mathphys-num]{sn-jnl}
\DeclareUnicodeCharacter{2212}{\textminus}

\usepackage{pgf}
\usepackage{layouts}
\usepackage{geometry}
\usepackage[acronym]{glossaries}
\usepackage{optidef}
\usepackage{soul}
\usepackage{color}
\usepackage{tabularx}
\usepackage{graphicx}%
\usepackage{multirow}%
\usepackage{amsmath,amssymb,amsfonts}%
\usepackage{amsthm}%
\usepackage{mathrsfs}%
\usepackage{pdflscape}
 \usepackage{svg}

\theoremstyle{thmstyleone}%
\theoremstyle{thmstyletwo}%

\theoremstyle{thmstylethree}%

\begin{document}

\title{Towards Privacy-Preserving Anomaly-Based Intrusion Detection in Energy Communities}

\author*[1]{\fnm{Zeeshan} \sur{Afzal}}\email{zeeshan.afzal@liu.se}
\author[2]{\fnm{Giovanni} \sur{Gaggero,}}
\author[1]{\fnm{Mikael} \sur{Asplund}}


\affil[1]{Department of Computer and Information Science (IDA) \\ Linköping University, 581 83, Linköping, Sweden}
\affil[2]{Department of Electrical, Electronic and Telecommunications Engineering, and Naval Architecture (DITEN) \\ University of Genoa, Via all'Opera Pia 11A, 16145, Genoa, Italy}


\abstract{


Energy communities consist of decentralized energy production, storage, consumption, and distribution and are gaining traction in modern power systems. However, these communities may increase the vulnerability of the grid to cyber threats. We propose an anomaly-based intrusion detection system to enhance the security of energy communities. The system leverages deep autoencoders to detect deviations from normal operational patterns in order to identify anomalies induced by malicious activities and attacks. Operational data for training and evaluation are derived from a Simulink model of an energy community. The results show that the autoencoder-based intrusion detection system achieves good detection performance across multiple attack scenarios. We also demonstrate potential for real-world application of the system by training a federated model that enables distributed intrusion detection while preserving data privacy.

} 


\keywords{smart grid, distributed energy resources, battery energy storage systems, cloud control systems, energy communities, cyber security}

\maketitle

\section{Introduction}

The increasing electricity demand across the world, coupled with the efforts towards a sustainable and zero-emission energy system, are significantly transforming the energy landscape. Distributed and renewable energy sources (RESs) are at the forefront of this transformation, providing a cost-effective, environment-friendly way to generate energy. However, compared to the reliable but fossil-fuel-based traditional resources, the power produced by RESs tends to be highly variable and unpredictable due to heavy dependence on weather conditions, such as sunlight and wind~\cite{Oshnoei_2020, Saini_2023, afzal2024security, mohan2020comprehensive, Ibraheem_2023}. For the electricity grid to operate, there must be an almost perfect balance between power production and consumption~\cite{Rinaldi_2022}. The consequences of large imbalances could be serious with an impact on safety, such as blackouts in given areas or damage to connected equipment which could result in fires. 

As distributed RESs become common and constitute a larger share of the energy mix, balancing supply and demand within the power grid to maintain a reliable energy supply becomes increasingly difficult~\cite{iva_svängmassa,problem_renewable_load_demand}. One solution to counteract the balancing difficulties in the grid is to use flexible assets such as battery energy storage systems (BESSs), heat pumps, electric vehicles (EVs), and vehicle-to-grid (V2G) systems, that can either store surplus power, quickly adapt their energy consumption, or inject more energy into the grid in the event of unforeseen supply and demand~\cite{flex_report}. 

Energy communities (ECs) is an initiative led by the European Union to accelerate the transition towards a cleaner, sustainable, and resilient energy future. In energy communities, citizens collectively generate, consume, store, share, and sell power from RESs to achieve common goals such as energy independence, environmental sustainability, and financial efficiency. At the heart of these energy communities are cloud-based management platforms that centrally manage these communities and the operation of flexible assets, including controllable loads, generators, batteries, and inverters, which are physically located within individual households. To optimize community and operations, these management platforms coordinate critical decisions, such as scheduling energy consumption, determining the timing and quantity of energy generation, and managing battery storage and discharge. 

Unfortunately, as previous research has shown~\cite{novel-EC, impact-EC}, ECs are vulnerable and increase the power grid's exposure to cyber threats. One reason for the increased risk is the cloud-based community management that typically relies on public communication networks, such as the Internet. Additionally, these management platforms also typically host a web service to allow community members to connect and access a dashboard to visualize their community metrics. This functionality enables many web application and authentication vulnerabilities and in combination with the outdated, yet prevalent usage of legacy protocols, expands the attack surface. Consequently, adversaries can remotely connect to the management platform and if they can infiltrate the platform by exploiting any vulnerability, they can gain unauthorized access and control over multiple assets such as generators, batteries, and flexible loads within the managed communities. They can then manipulate the operation of assets to disturb the delicate balance between energy generation, storage, and consumption. These vulnerabilities make the ECs vulnerable to many potential attacks, which have recently been shown to be practical~\cite{security-bess} and potentially devastating for the electrical grid. The situation is critical as the quantity and capacity of flexibility in the grid is expected to significantly increase over the coming years.


Securing ECs is essential to maintaining the integrity and availability of the electrical grid. Therefore, detection and mitigation of potential attacks exploiting the vulnerabilities is of paramount importance. However, existing detection methods are not well-suited for energy communities, as they do not consider the complex interdependencies between generation, storage, and consumption. To address this, we propose an anomaly-based intrusion detection system (IDS) specifically designed for energy communities. The IDS is proposed to be deployed within each household and connected to the local area network (LAN) on which assets like generators with inverters, batteries, and flexible loads are connected. The goal of the proposed system is to detect anomalies in power generation, storage, and consumption that may indicate an attack. By detecting such threats, the IDS aims to ensure the continued stability and security of ECs, facilitating their safe and sustainable integration into the modern smart grid. The key contributions of this work are as follows:

\begin{enumerate}

     \item an open-source operational dataset for an EC, including both benign and malicious scenarios. 
    \item an IDS framework capable of detecting anomalies in energy communities.
    \item demonstration of real-world potential of the framework by using a federated model that enables detection while preserving privacy.

    \end{enumerate}

The remainder of this paper is structured as follows. Section 2 provides the necessary background and an overview of the related work. It also highlights the gaps in the literature that we aim to fill. Section 3 describes our method; the simulation environment, attack scenarios, autoencoder-based intrusion detection and its formalization, and an approach to enable a privacy-friendly detection using federated learning. Section 4 presents the experimental results and evaluates the proposed system's detection performance. Section 5 provides a discussion on the results, their implications, and outlines future research directions. Finally, section 5 provides conclusions.

\section{Background and Related Work}
\label{section:related_work}

The evolution of traditional power grids into smart grids represents a significant advancement in the management and distribution of electric power. However, the increased connectivity and reliance on digital technologies also introduce new security vulnerabilities. This makes smart grids a prime target for cyber attacks~\cite{flex-threat}. As the smart grid is a critical infrastructure, breaches can lead to severe consequences, including widespread power outages, financial losses, and even threats to national security~\cite{iva_svängmassa}. The increasing trend and practice of monitoring and managing distributed energy resources (DERs) and other smart grid assets remotely through cloud-based solutions over the Internet further complicates the security landscape~\cite{Trevizan_2022, Naseri_2023, Kharlamova_2020}.

Several studies~\cite{mohan2020comprehensive, mohammad2023, smartgrid-dos} have investigated potential attacks in smart grids and revealed attacks such as time delay switch (TDS) attacks, false data injection (FDI) attacks, denial of service (DoS), and replay attacks. Mahmoud et al.~\cite{mahmoud2019modeling} showed that DoS attacks can cause instability of power grids and produce lengthy delays between packets being sent and received. The potential risks associated with the control of a high number of DERs across the distribution grid are also studied in the literature~\cite{dabrowski_load_altering}. For instance, load altering attacks~\cite{dabrowski_load_altering} can be launched by an attacker that physically or remotely controls a large enough part of the load (power) in an area or by somehow manipulating aggregated load communication of demand aggregation systems~\cite{mohan2020comprehensive}. Such large-scale and distributed attacks against the power grid are considered among the most critical threats~\cite{Trevizan_2022}. If an attacker gets access to enough flexible assets, such as electric vehicle (EV) chargers, they could coordinate consumption patterns that might impact the stability of the grid. Another study has focused on the poor encryption in the communication between BESSs installed in homes and the manufacturer servers used for remote control~\cite{Baumgart_2019}, leading to threats. 

\subsection{Energy Communities and their Security}

Although, the electric grid congestion and imbalance challenges are long known, these problems are lately approached through the perspective of energy flexibility. Flexibility is the ability to adjust power generation or demand to account for grid conditions. Many actors in the grid together contribute to the flexibility of the grid. For example, flexibility is usually offered by electricity consumers or prosumers that own generation and storage facilities and controllable loads. To increase energy flexibility and with the advancement of regulations regarding  reduction in carbon emissions, the European Union published the renewable energy directive (REDII)~\cite{ec-eu} and introduced mechanisms such as energy communities (ECs). There are many definitions for ECs, but the idea with the emergence of ECs is local energy production from distributed and RESs and its storage and sharing among members in the proximity of the production facility to reach common goals such energy independence, environmental or economic. As ECs introduce unique challenges due to their distributed nature and cloud-based management, which could be manipulated at large scale to manipulate energy demand or inject extra energy into grid, it is critical to study and address their security implications. 

To identify existing research literature on security for ECs, we conducted a comprehensive review of the existing literature. The boolean search string (``energy communities” AND *security) is used to search relevant databases for peer-reviewed research including, journal, conference, and workshop papers as well as peer-reviewed book chapters. The search is focused on IEEE Xplore digital library, ACM digital library, Scopus, and Web of Science databases. We excluded matches before the year 2018 and non-peer reviewed abstracts and publications from the search. The string is searched in the title, abstract, and keywords of the article. Table~\ref{tab:papers} shows a summary of identified literature per database. 

\begin{table}[htbp!] 
\centering
\caption{Identified papers per source.}
\begin{tabular}{| p{0.2\linewidth} | p{0.2\linewidth} |}
\hline
\textbf{Database} & \textbf{Papers} 
\\ \hline
IEEE & 43 \\ \hline
ACM & 44 \\ \hline
Scopus & 91 \\ \hline
Web of Science & 51 \\ \hline
\end{tabular}
\centering
\label{tab:papers}
\end{table}


Our initial search resulted in a total of 229 papers, as shown in Figure (step 0). In the next step, we removed duplicates and applied the following two exclusion criteria (ECs) on the remaining 162 papers; EC1: the publication is not concerned with energy communities, and EC2: the publication does not relate to any cybersecurity or privacy aspect. In step 2, we read the paper title and abstract to identify 5 relevant papers in step 3.

\begin{figure}
    \centering
    \includegraphics[width=1\linewidth, height=6cm]{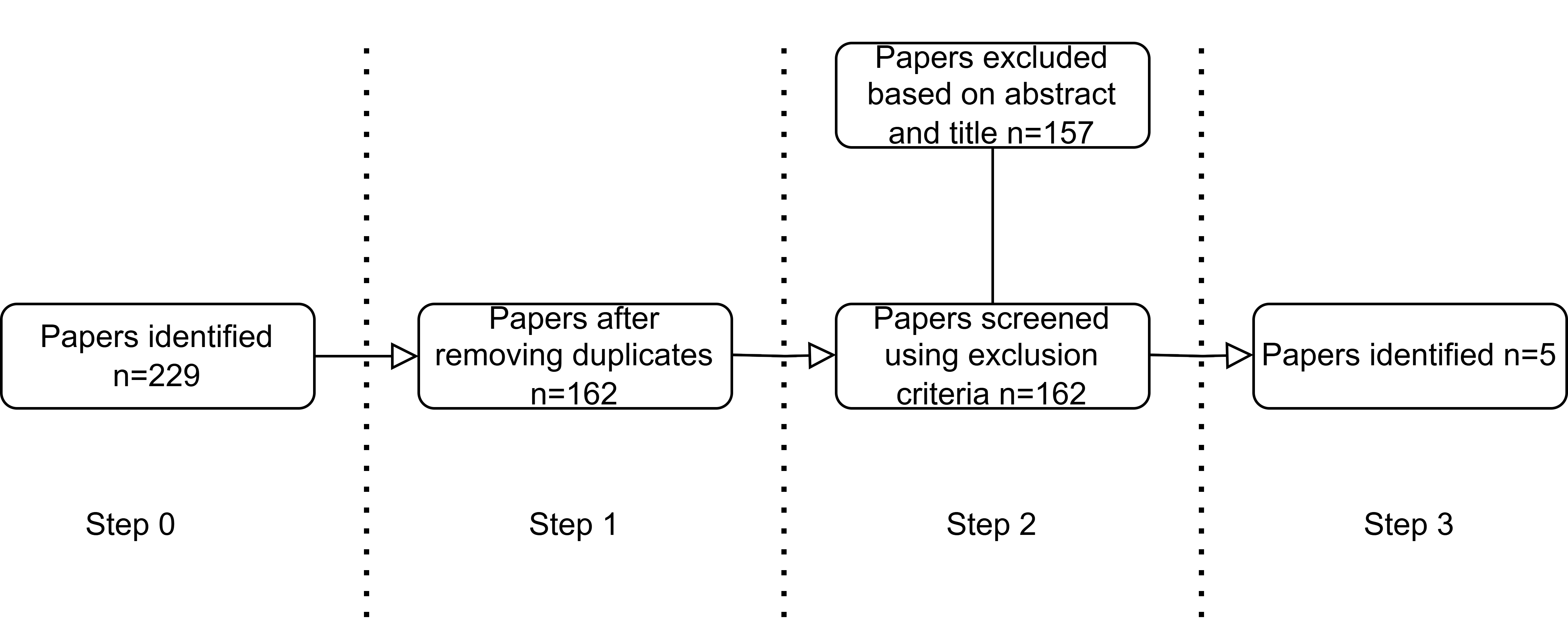}
    \caption{Literature review process.}
    \label{fig:enter-label}
\end{figure}

From the final selection of papers, only a handful of papers actually address the cybersecurity aspects of ECs. Gaggero et al.~\cite{novel-EC} analyze the potential architectures and protocols used to build ECs and evaluated possible vulnerabilities and attack vectors. Three main attack vectors were identified; attacks against communication protocols, attacks against smart gateway, and attacks against the platform. The paper also discussed some solutions which can be employed to mitigate the risk. In another work, Mokarim et al.~\cite{impact-EC} focus on the software platforms responsible for managing and controlling ECs as they handle a lot of sensitive information. The presented attack model in the research work assumes that an attacker can gain full control over such a software platform using common attacks on servers and applications and then disrupt the operation of the distribution grid by altering certain parameters (set points) of the system. The analysis is done by observing how the grid responds to these power manipulations for both low-voltage (LV) as well as medium-voltage (MV) systems. The power manipulations led to either absorption or injection of power in the grid. The paper highlights that renewable energy communities must implement proper cybersecurity monitoring tools and DSOs must maintain the ability to disconnect generators and loads. Another work~\cite{should-we} proposes to implement automatic mechanisms in response to cyber attacks in ECs. As the generators and inverters exist in the homes of private individuals, this makes it impossible to have dedicated cybersecurity teams that implement a proper incident response. The proposed approach includes connecting an IDS to every local generator/inverter. In the event of detected attack/s, the IDS signals electrical protection devices such as a circuit breaker to disconnect the particular generator under attack from the grid. The study~\cite{micro-ec} proposes to model a microgrid energy community, in order to simulate different cyberattacks to assess their spread and impact and finally develop methods to protect from those attacks. The considered attacks include FDI where an attacker modifies the measured information neighboring agents by adding false data and hijack attack where the attacker replaces measurements with malicious data. The attacks are simulated on a customized IEEE34 test feeder and results show deteriorated quality of voltage regulation by the affected inverters. 

Sanduleac et al.~\cite{ec-resilience} focus on the resilience of an EC. Resilience is seen both as energy resilience and as a cybersecurity aspect against attacks. The work views ECs as citadels that need to build virtual walls to be resilient against any external aggression. This entails that energy communities need to be always operational and sustainable in face of harsh external conditions such as Internet blackout or cyberattacks. The paper presents a communication architecture that an EC can use to secure all its interactions over the Internet with the DSO and other market actors. As such digital interactions are sensitive and need clear rules, the notion of contracts is put forward which can be exchanged between actors. The concept has been implemented in a simulated energy community and the data exchange is further secured by common ICT security solutions such API keys, digital signatures, VPN, and HTTPS. Steinheimer et al.~\cite{ec-p2p1} present different approaches for a service management framework to control and monitor decentralized energy consumers, storages, and generators. In these approaches the user is integrated in design and configuration of the services for energy management which offer the possibility to follow its personal needs. ECs between  households are discussed as one of the approaches and P2P interaction among the members is shown to offer a solution to solve user rights and privacy restrictions and to achieve common goals. Other literature works focus on case-studies and pilot projects~\cite{ec-italian, ec-dutch}. Additionally, they focus on energy security rather than cybersecurity~\cite{ec-energysecurity}, explore policy implications and legals aspects for ECs~\cite{ec-legal1, ec-legal2}, and examine the application of distributed ledger technologies, such as block chain, in the context of ECs~\cite{ec-dlt1, ec-dlt2, ec-dlt3}.



\subsection{Intrusion Detection in Smart Grids} 
Intrusion detection systems (IDSs) can play a critical role in protecting smart grids and renewable energy systems from cyber-physical threats. Over the years, researchers have studied various rule-based and signature-based techniques to detect threats exploiting vulnerabilities associated with DERs and their communication networks~\cite{ids-compilation1, ids-compilation2, ids-rule}. However, these methods are limited in their ability to identify zero-day attacks or novel intrusion strategies, which are increasingly common in smart grid environments. To address this limitation, anomaly-based IDS methods have gained significant attention~\cite{ids-ml}. These systems analyze operational data to establish a baseline of normal behavior and detect deviations that may indicate malicious activity. Several studies have demonstrated the effectiveness of anomaly-based IDS in identifying a wide range of cyber-physical attacks~\cite{ids-ml2}. However, the heterogeneity and dynamic nature of ECs introduce unique challenges that require more sophisticated detection mechanisms.

Autoencoders are a type of deep artificial neural network designed for unsupervised machine learning~\cite{autoencoders}. They work by learning a compressed representation of normal data and reconstructing it with minimal loss. Anomalies, which deviate from the learned representation, result in higher reconstruction errors, making them detectable. In the context of smart grids, several studies~\cite{ids-auto1,ids-auto2,ids-auto3} have utilized autoencoders to identify anomalies in sensor data, control signals, and energy flows. For example, Takiddin et al.~\cite{ids-auto1} propose a detection method for electricity theft in smart grids using deep autoencoders. This approach captures complex consumption patterns and temporal correlations in energy usage data. Simulation results demonstrate that the proposed method enhances detection rate and reduces false alarms as compared to existing solutions. Similarly, Nazir et al.~\cite{ids-auto4} employ an autoencoder-based IDS to detect FDI attacks in SCADA systems, achieving high detection accuracy while maintaining low false-positive rates. 

\subsection{Gaps and Opportunities}

While significant progress has been made in the development of intrusion detection mechanisms for smart grids in general, several gaps remain. Existing anomaly detection methods do not consider the complex interdependencies between generation, storage, and consumption. As demonstrated by our literature review, security aspects in ECs are not thoroughly studied. The challenges posed by the integration of cloud-based management platforms to these ECs demand research in this direction. Finally, despite the potential of autoencoders for anomaly detection, their application in ECs to detect a wide range of cyber-physical attacks requires thorough investigation.

This work addresses these gaps by proposing an autoencoder-based IDS tailored specifically to the unique characteristics of ECs. By leveraging unsupervised learning techniques, the system can detect multiple attacks. Furthermore, the proposed IDS is designed to operate within the local EC infrastructure, facilitating real-time detection and response to cyber-physical threats. This paper builds on the insights and methodologies from previous research while extending their application to ECs, thereby contributing to the growing body of knowledge on smart grid security.
\section{Privacy-Preserving Intrusion Detection} 
\label{section:approach}

This section provides an overview of our approach towards a privacy-preserving anomaly-based IDS. It describes the proposed data-driven detection model together with problem formalization, and an approach to enable privacy-friendly detection using federated learning.

\subsection{Autoencoder-based Intrusion Detection}

Autoencoders have emerged as an effective tool for anomaly detection, particularly in cyber-physical systems. By learning compressed representations of normal system behavior, these models are efficient at reconstructing input data with lower error under normal conditions. Anomalies, which deviate significantly from learned patterns, result in higher reconstruction errors, enabling their identification. This data-driven approach is particularly advantageous for cyber-physical systems, where anomalies are often rare, and explicit labeling is impractical. In detail, the application of autoencoders involves the following key steps:

\begin{enumerate}
    \item Training: The autoencoder is trained using data representing the normal operation of the system. During training, the network learns to encode input data into a lower-dimensional latent space and reconstruct it back to its original form. The goal is to minimize the reconstruction error.
    \item Reconstruction Error: Once trained, the autoencoder can process new data and attempt to reconstruct it. For normal data, reconstruction errors are typically small, as the autoencoder has learned these patterns during training. However, for anomalous data, which deviates significantly from normal patterns, reconstruction errors tend to be larger.
    \item Setting a Threshold: To distinguish between normal and anomalous behavior, a reconstruction error threshold is determined. This threshold is often set based on statistical measures (e.g., mean and standard deviation of reconstruction errors on the training set) or by analyzing the error distribution during validation.
    \item  Testing and Detection: During testing, the autoencoder processes new data and computes the reconstruction error for each sample. If the error exceeds the predefined threshold, the data point is flagged as an anomaly. This enables both real-time or offline anomaly detection in dynamic systems.
\end{enumerate}

\subsection{Problem Definition}

We focus on detecting anomalies in the EC’s operational data using a time-series anomaly detection model based on a reconstruction autoencoder model. The problem is formalized as follows. Let the energy community's operational data be represented as a multivariate time series:

\begin{equation}
\mathcal{X} = \{X_1, X_2, \dots, X_T\},
\end{equation}

where each $X_t \in \mathbb{R}^d$ is a feature vector of $d$ dimensions at time step $t$, meaning there are $d$ different numerical measurements recorded at each time step. Each feature represents an energy or power related measurement, such as battery current ($I_{\text{batt}}$), load power ($P_{\text{load}}$), and voltage ($V$). The goal is to distinguish between normal operational data and anomalous data based on probability distributions. The model assumes two types of data distributions, as described below.

\begin{itemize} 
    \item \textbf{Normal Data:} When the energy system operates under normal conditions, its data follows a probability distribution:
    \begin{equation}
    X_t \sim p_{\text{normal}}(X).
    \end{equation}

    \item \textbf{Anomalous Data:} Attacks cause deviations from $p_{\text{normal}}(X)$, generating out-of-distribution samples:
    \begin{equation}
    X_t \sim p_{\text{attack}}(X), \quad \text{where} \quad p_{\text{attack}}(X) \neq p_{\text{normal}}(X).
    \end{equation}

An autoencoder $f_\theta: \mathbb{R}^{d \times T} \to \mathbb{R}^{d \times T}$ where d x T represents operating on an entire segment of length T, learns to reconstruct normal data. The goal is to train the autoencoder to learn an identity mapping:
    \begin{equation}
        f_\theta(X) \approx X.
    \end{equation}
    where \( f_\theta(X) \) is the autoencoder function parametrized by \( \theta \).  Given an input $X$, the model reconstructs it as:
    \begin{equation}
        \hat{X} = f_\theta(X).
    \end{equation}

The model is trained only on normal data, so it becomes good at reconstructing normal patterns but struggles with anomalous data. In other words, it has low reconstruction error for normal samples and high reconstruction error for anomalies. The reconstruction error $e$ is computed using the \textit{Mean Absolute Error (MAE)}:
    \begin{equation}
        e(X) = \frac{1}{dT} \sum_{i=1}^{d} \sum_{t=1}^{T} |X_{t, i} - \hat{X}_{t, i}|.
    \end{equation}
This calculates the average difference between the original and reconstructed data. Higher error means the model failed to reconstruct the input well, indicating a potential anomaly. For anomaly classification, a threshold $\tau$ is determined from the reconstruction error $e$ distribution of normal data. A given sample $X$ is thus classified as anomalous if:
    \begin{equation}
        e(X) > \tau.
    \end{equation}
    Otherwise, it is considered normal.

\end{itemize}

\subsection{Private Anomaly Detection}

In traditional and centralized machine learning, data from different sources is aggregated into a single dataset and transferred to a central server, where the model is trained. Since raw data is required to be sent to a central location, this approach suffers from several drawbacks. First, there are privacy risks with collecting and transferring raw sensitive data. Second, this approach requires high network bandwidth, and third, the central server could prove to be a single point of failure. Federated learning (FL) on the other hand is a decentralized machine learning approach which allows multiple devices or nodes to collaboratively train a shared model without sharing their local raw data. Instead of transferring data to a single central server, federated learning allows each node to train the model on its local data and share only the updated model parameters. This approach preserves data privacy as instead of data moving towards the model, the model is essentially moved closer to the data, reduces communication costs, and is particularly valuable in scenarios where data is sensitive or distributed across various locations, such as in ECs.


The FL process begins with the server initializing a global model and distributing its parameters to the clients. Each client trains the model on its local dataset, which consists of normal data, and computes updates to the model parameters. These updates are then sent back to the server, which is responsible for aggregation to improve the global model and distribution of the global model parameters to the clients. This process repeats for a predefined number of communication rounds until the training process is complete, and the trained global model is distributed to all the clients.

In ECs, where multiple households or facilities generate and consume energy, detecting anomalies is critical for ensuring reliable and efficient grid operations. By combining FL with autoencoders, it is possible to build a robust, privacy-preserving anomaly detection system tailored to such communities. Each node in the EC (e.g., a household or facility) can train an autoencoder locally using its energy consumption and generation data. The autoencoder learns to reconstruct normal patterns specific to the node. When new data deviates significantly, resulting in high reconstruction errors, anomalies can be detected locally. This combination offers several advantages: it preserves the privacy of individual nodes by keeping their data local, accommodates the heterogeneity of energy patterns across the community, and ensures a more accurate and adaptive anomaly detection system through collaborative learning.



\section{Evaluation}

This section provides an evaluation of the proposed IDS. We begin by introducing the simulation environment used to model an EC and collect data for assessment. Next, we outline various attack scenarios, followed by a description of the model's architecture, its parameters, and the feature vector. Finally, a detailed assessment of the detection performance for the investigated attacks is presented.

\subsection{Simulation Environment}

\begin{figure}[htbp!] 
\centering
\includegraphics[width=25cm,height=12cm,keepaspectratio, angle=90]{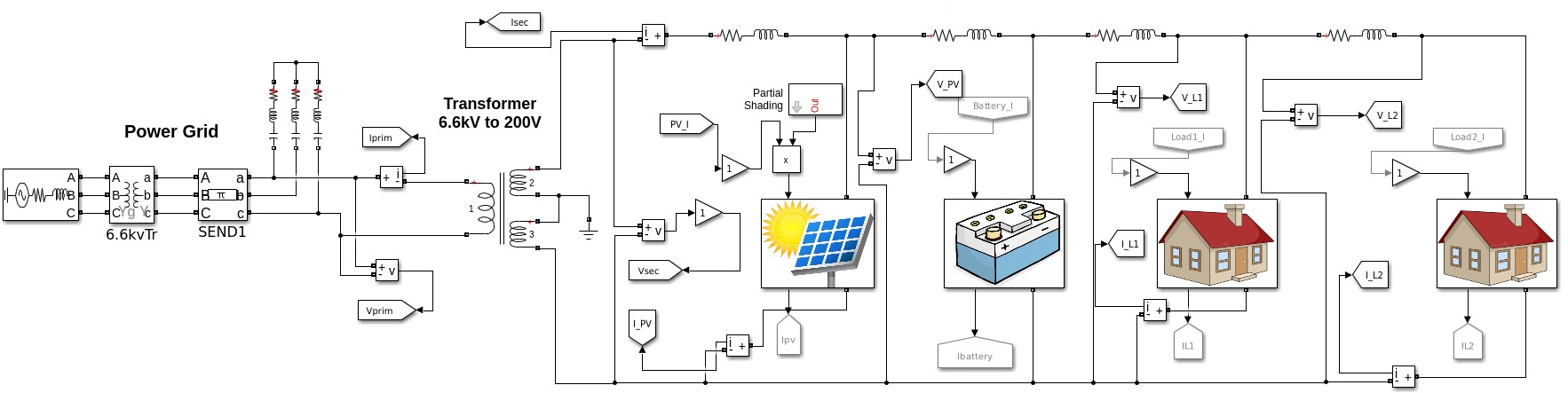}
\caption{Simulink model for a small EC.}
\label{fig:model}
\end{figure}

We employ simulations to gather data for training and testing our proposed detection system as it is not possible to access data from live and real-world systems due to data privacy and other regulations. The model is implemented in Matlab/Simulink  and builds on top of the model available at~\cite{ec-model}. It illustrates the behavior of a small-scale EC over a 24-hour period. An overall scheme is shown in Figure~\ref{fig:model}. The model represents an EC designed to simulate energy generation, storage, consumption, and control. It is connected to single-phase electrical grid that serves as the primary power source via a transformer, which reduces the voltage from 6.6 kV to 200 V. The EC has two local generators: a solar power generation system and a storage battery. The solar power system generates renewable energy with output varying depending on sunlight, peaking at 5 kW between 14:00 and 15:00 while producing no power from 20:00 to 04:00. To represent variable generation due to weather conditions such a cloud cover, the solar system only generates lower power from 11.00 to 12.00. The storage battery acts as a buffer, storing excess power when generation surpasses consumption and supplying power during shortages. The battery is managed by a battery controller, which ensures efficient power distribution. The controller ensures that the batter absorbs surplus power when energy generation exceeds the EC's needs and provides additional power during shortages. Two households with controllable and flexible loads inside are connected to the EC. Each household consumes a maximum of 2.5 kW of power with variations throughout the day based on typical consumption patterns. Peak consumption occurs at 09:00 (around 4 kW), as well as at 19:00 and 22:00 (around 5 kW). Both the solar power system and storage battery, being DC power sources, rely on converters to produce single-phase current suitable for the grid.

The control strategy for the EC is designed to minimize dependence on the main power grid, ensuring that the combined power from solar generation and the storage battery meets the households' demands whenever possible. The battery controller operates during specific periods to achieve this. From 00:00 to 12:00 and again from 18:00 to 24:00, the battery controller actively manages the battery's charge and discharge to ensure that the active power flowing into the system from the transformer is approximately zero. During this period, the storage battery supplies power during shortages and absorbs excess energy when available.

\begin{figure}[htbp!] 
    \centering 
    \includegraphics[width=1.0\textwidth]{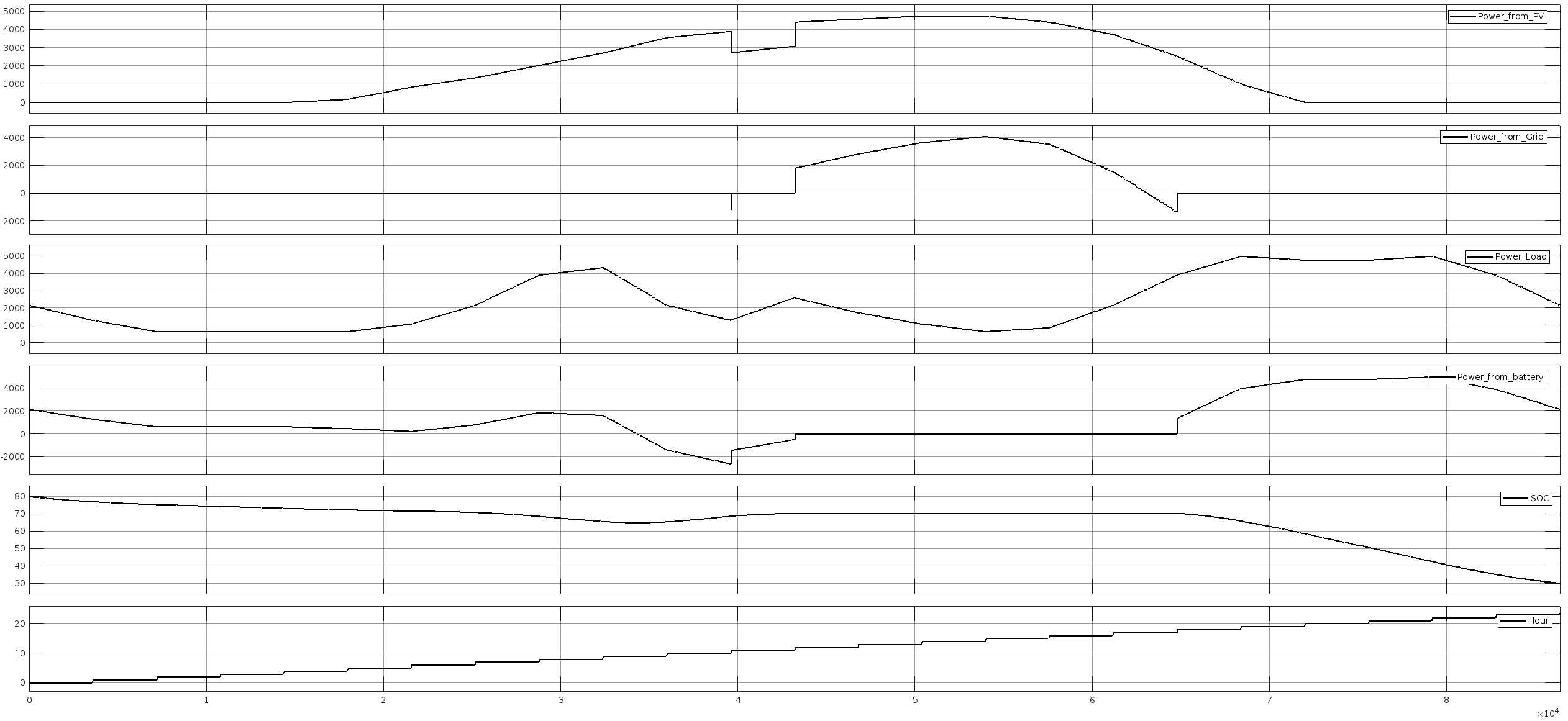}
    \caption{Plots for normal operation of the EC.}
    \centering
    \label{fig:ec-plot}
\end{figure}

Between 12:00 and 18:00, the battery controller is inactive. During this time, the battery’s State of Charge (SOC) remains constant as it neither charges nor discharges. Any power shortages during this period are supplied by the grid, while any surplus power is fed back. In summary, the households in the model receive electrical power as their primary input to meet their energy demands. The output from the households is the electrical load they present to the EC. The model effectively demonstrates the behavior of a small-scale EC, showcasing how it balances energy generation, consumption, and storage while maintaining reliable operation. Figure~\ref{fig:ec-plot} shows plots for normal operation of the EC. 

\subsection{Attack Scenarios}

The aim of the proposed IDS is to detect anomalies that indicate disruption or manipulation of the normal operation of an EC as described above. In this context, we are interested in detecting the effects of attacks rather than their actual mechanisms. The focus is on detecting threats that can manipulate power generation, its storage, and the demand of flexible loads or assets. We consider that the adversaries can exploit vulnerabilities to launch many common attacks including FDI and DoS attacks to control and manipulate the growing number of remotely controlled flexible assets in ECs, potentially affecting stability of the power grid. For instance, an adversary can intercept communication data or signals between assets and manipulate them, in order to control what the receiver sees. In the context of an EC, an adversary can modify the control signals exchanged between the community manager and generators such as storage batteries. By injecting malicious data, power set-points could be modified allowing the attackers to mislead assets into making decisions based on falsified information, potentially leading to incorrect decisions and resulting in attacks such as power injection or absorption from the grid.

On the other hand, adversaries can also target the demand side to launch load altering attacks (LA). These attacks manipulate the demand of remotely controlled flexible loads such as electrical heaters and other smart appliances that have the potential to consume large amounts of energy, in order to disrupt the balance between power supply and demand. Increasing or decreasing power consumption of a single asset may not be critical in isolation, but if an attacker can manipulate demand of multiple assets, which is a real possibility in ECs, grid stability could be in danger. The following four attack scenarios form the basis of our investigation.

\begin{itemize}
    \item \textbf{Power Injection (PI)}: An attacker increases the power supplied to the electrical grid by unexpectedly discharging controlled flexible assets, such as batteries.
    \item \textbf{Power Absorption (PA)}: An attacker decreases the power supplied to the electrical grid by unexpectedly charging controlled flexible assets, such as batteries.
    \item \textbf{Load Increase (LI)}: An attacker increases the energy demand of controlled assets, resulting in an imbalance between generation and consumption.
    \item \textbf{Load Reduction (LR)}: An attacker decreases the energy demand of controlled resources, resulting in an imbalance between generation and consumption.
\end{itemize}

\subsection{Detection Model Architecture}




We implement an autoencoder using Long Short-Term Memory (LSTM) networks for time series anomaly detection in Python using Keras API. The model consists of two main parts: an \textbf{encoder} and a \textbf{decoder}.  

\subsection*{Encoder}  
The encoder is composed of two stacked LSTM layers:  
\begin{itemize}  
    \item The first LSTM layer has \textit{128 units} and processes input sequences of length \textit{TIME\_STEPS} (10 in our case). The feature dimension corresponds to the number of variables in the dataset. We set \texttt{return\_sequences=True} to preserve temporal dependencies in the data.  
    \item The second LSTM layer has \textit{64 units} and does not return sequences, meaning it compresses the input into a lower-dimensional latent representation.  
\end{itemize}  

\subsection*{Decoder}  
To reconstruct the original sequence, the decoder follows these steps:  
\begin{enumerate}  
    \item A \textit{RepeatVector} layer replicates the latent representation across the required time steps.  
    \item Two LSTM layers mirror the encoder’s structure but in reverse:  
    \begin{itemize}  
        \item A \textit{64-unit LSTM layer} that returns sequences.  
        \item A \textit{128-unit LSTM layer} that also returns sequences.  
    \end{itemize}  
    \item Finally, a \textit{TimeDistributed Dense layer} with a linear activation function generates the output sequence, ensuring each time step has the same dimensionality as the input.  
\end{enumerate}

\begin{figure}[htbp!] 
    \centering 
    \includegraphics[width=1.0\textwidth]{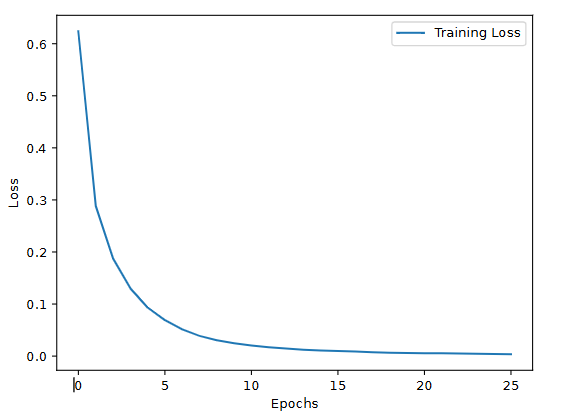}
    \caption{Training Loss}
    \centering
    \label{fig:training1}
\end{figure}

The model is trained using the \textit{Adam} optimizer with a \textit{mean squared error} (MSE) loss function, optimizing its ability to reconstruct input sequences. Based on the observed losses over epochs during training, as shown in Figure~\ref{fig:training1}, hyperparameters are determined. Epochs are set to 50 with a batch size of 128, using a 90-10\% train-validation split. An early stopping mechanism is implemented to prevent overfitting, monitoring the validation loss and halting training if no improvement is observed for five consecutive epochs. Anomalies are detected based on reconstruction error, calculated as the mean absolute error (MAE) between input and reconstructed sequences. Since the model processes time series sequences in a sequential manner, the total count of trainable parameters, including all LSTM weight matrices, bias terms, and the dense layer, results in 256,270 trainable parameters.




\subsection{Data}

Table~\ref{tab:data} represents the measures extracted from the simulated model. Together, these measures form the feature vector $X_t$ extracted at time $t$ and are used in training and testing of the detection model. The dataset used for training consists of simulating the EC for 24 hours of typical (normal) operating conditions. 
However, as LSTM autoencoders can benefit from learning how features evolve over time leveraging their memory capabilities to differentiate between normal variations and genuine anomalies, we create sequences of input data to capture the temporal dependencies in the data. The data is structured into overlapping sequences of 10 time steps.
 The test dataset is separately collected and contains time series data structured into sequences of 10 time steps.


\begin{table}[htbp!] 
\centering
\caption{Features used for training and detection.}
\begin{tabular}{| p{0.2\linewidth} | p{0.2\linewidth} | p{0.2\linewidth} |}
\hline
\textbf{Feature vector} & \textbf{Measurement} & \textbf{Description}
\\ \hline
$X_1$ & $V_{\text{sec}}$ & Voltage secondary
\\ \hline
$X_2$ & $V_{\text{PV}}$ & Voltage solar
\\ \hline
$X_3$ & $V_{\text{L1}}$ & Voltage load 1
\\ \hline
$X_4$ & $V_{\text{L2}}$ & Voltage load 2
\\ \hline
$X_5$ & $I_{\text{sec}}$ & Current secondary
\\ \hline
$X_6$ & $I_{\text{PV}}$ & Current solar
\\ \hline
$X_7$ & $I_{\text{L1}}$ & Current load 1
\\ \hline
$X_8$ & $I_{\text{L2}}$ & Current load 2
\\ \hline
$X_9$ & $I_{\text{battery}}$ & Current battery
\\ \hline
$X_{\text{10}}$ & $P_{\text{PV}}$ & Power Solar
\\ \hline
$X_{\text{11}}$ & $P_{\text{L1}}$ & Power load 1
\\ \hline
$X_{\text{12}}$ & $P_{\text{L2}}$ & Power load 2
\\ \hline
$X_{\text{13}}$ & $battery_{\text{soc}}$ & Battery soc
\\ \hline
$X_{\text{14}}$ & $battery_{\text{ah}}$ & Batter Ah
\\ \hline

\end{tabular}
\centering
\label{tab:data}
\end{table}

\subsection{Anomaly Threshold}


\begin{figure}[htbp!] 
    \centering 
    \includegraphics[width=1.0\textwidth]{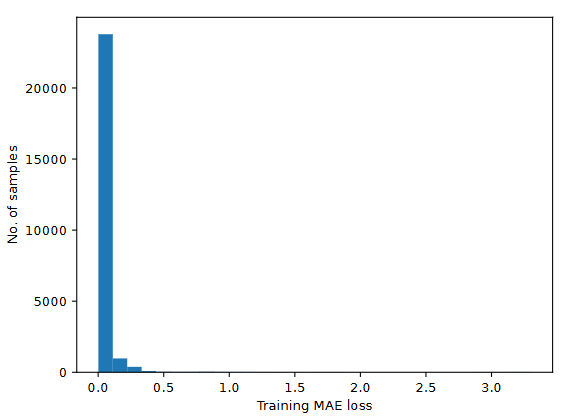}
    \caption{Distribution}
    \centering
    \label{fig:training2}
\end{figure}

After the training process, we study the reconstruction error over the training dataset. Figure~\ref{fig:training2} shows the distribution of reconstruction error over the training dataset. A threshold for anomaly detection can then be determined, and anomalies can be detected by comparing test reconstruction errors against this threshold to flag samples with errors exceeding the threshold. There are different ways to determine and optimize a threshold, depending upon the needs on precision and accuracy. We define threshold $\tau$ as the 95th percentile of the training loss. A sample $X$ is thus classified as anomalous if: \[ e(X) > \tau \quad \text{where} \quad \tau = \text{percentile}(e(X), 95) 
\]

\subsection{Results}
\label{results}

In the following subsection, an evaluation of the detection performance for the various cyberattacks discussed above is provided. To simulate attacks, we employ the gain block in Simulink which multiplies the given input by a constant value or gain. Thus, attacks are imitated by maliciously amplifying or curtailing the magnitude of the control signals. This results in many attack scenarios, for example, the battery may charge or discharge, regardless of the actual grid conditions and capacity, if the signal from controller to battery is compromised. The magnitude of the signal to the battery determines the charging rate of the battery, with a larger and positive amplitude telling the battery to charge faster (store energy) and a negative signal instructing the battery to discharge. Similarly, manipulation on the load side can result in an increase or decrease of the signals, triggering cascading events and incorrect controller actions. 

\subsubsection{Power Absorption Attacks}

\begin{figure}[htbp!] 
    \centering 
    \includegraphics[width=1\textwidth]{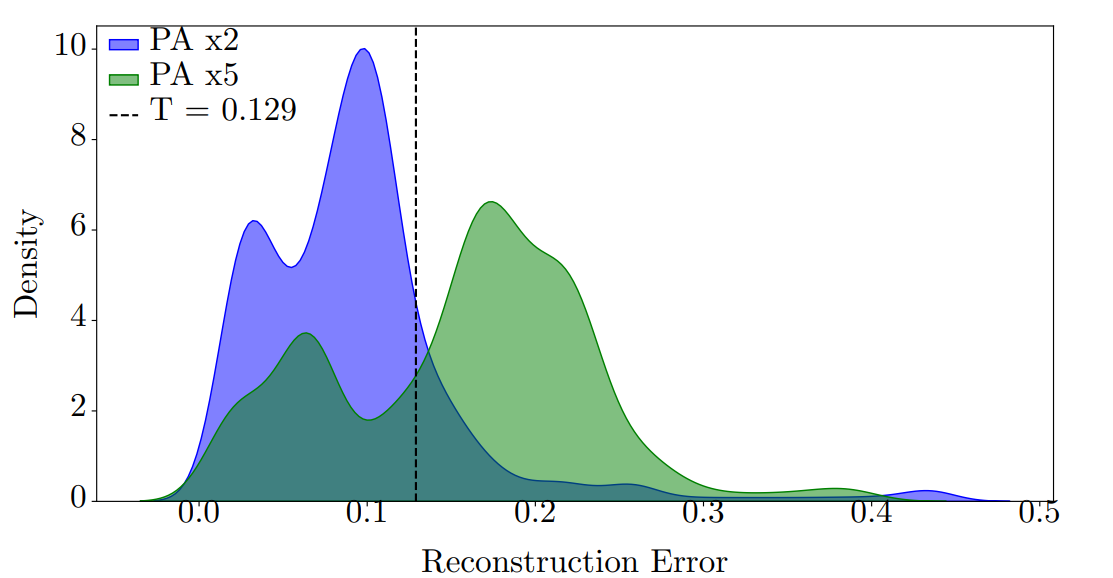}
    \caption{Distribution of Reconstruction error for PA attacks.}
    \centering
    \label{fig:PA1}
\end{figure}

\begin{figure}[htbp!] 
    \centering 
    \includegraphics[width=1.1\textwidth]{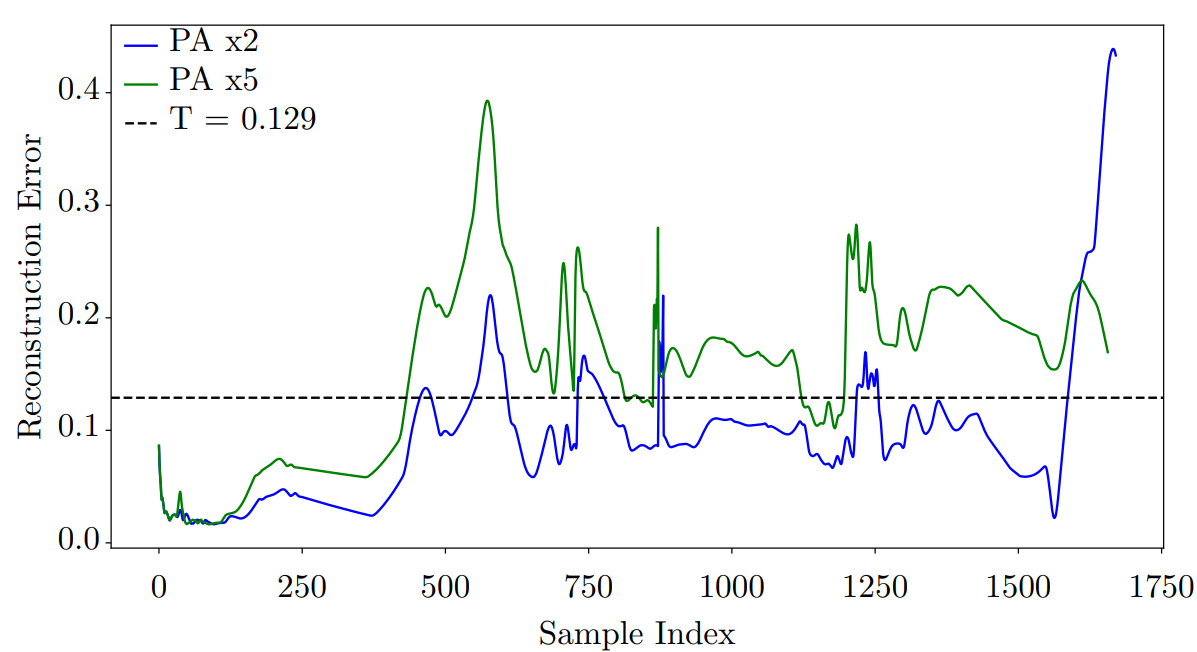}
    \caption{Detection performance for PA attacks.}
    \centering
    \label{fig:PA2}
\end{figure}

PA attacks force the battery to increase the charging rate and store more energy than expected. For this scenario, we experiment with two different variations. In the first instance, the control signal between controller and battery is doubled in amplitude to essentially tell the battery to charge at a rate that is twice than what is normal. Then, we further amplify the charging rate of the battery by multiplying the signal by five times resulting in a difference of 500\%. Figure~\ref{fig:PA1} shows the distribution of reconstruction error for these attack samples. The dotted line (T) indicates the detection threshold. Figure~\ref{fig:PA2} plots the detection performance of the model with attacks starting from sample 0.

It can be seen from  the plot that the model struggles to detect the attack samples when the amplitude is twice (PA x2) as normal, with a majority of attack samples achieving a lower reconstruction error than the threshold, resulting in them being classified as false negatives. However, the detection performance improves as the amplitude increases (PA x5) with the attack samples correctly classified as attacks by the model. We hypothesize that the detection model needs a bigger deviation from normality in this case to be more accurate in its detection. At the same time, although the model misses attack samples in the PA x2 scenario, it can still be useful as the attack can still be detected as long as there are large enough samples.

\begin{figure}[htbp!] 
    \centering 
    \includegraphics[width=1.0\textwidth]{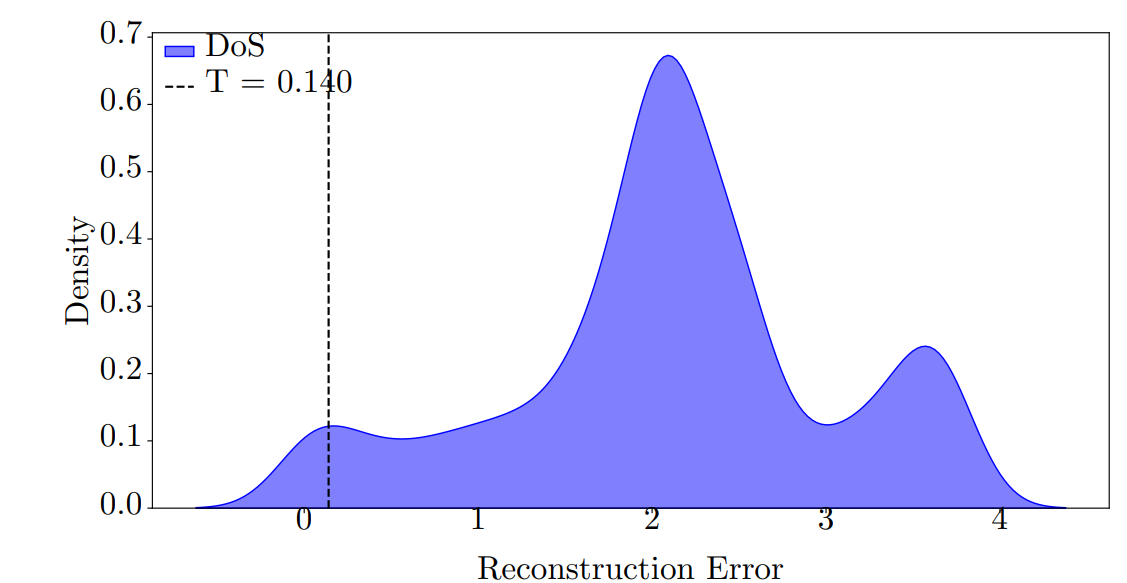}
    \caption{Distribution of Reconstruction error for DoS attack.}
    \centering
    \label{fig:dos1}
\end{figure}

\begin{figure}[htbp!] 
    \centering 
    \includegraphics[width=1.1\textwidth]{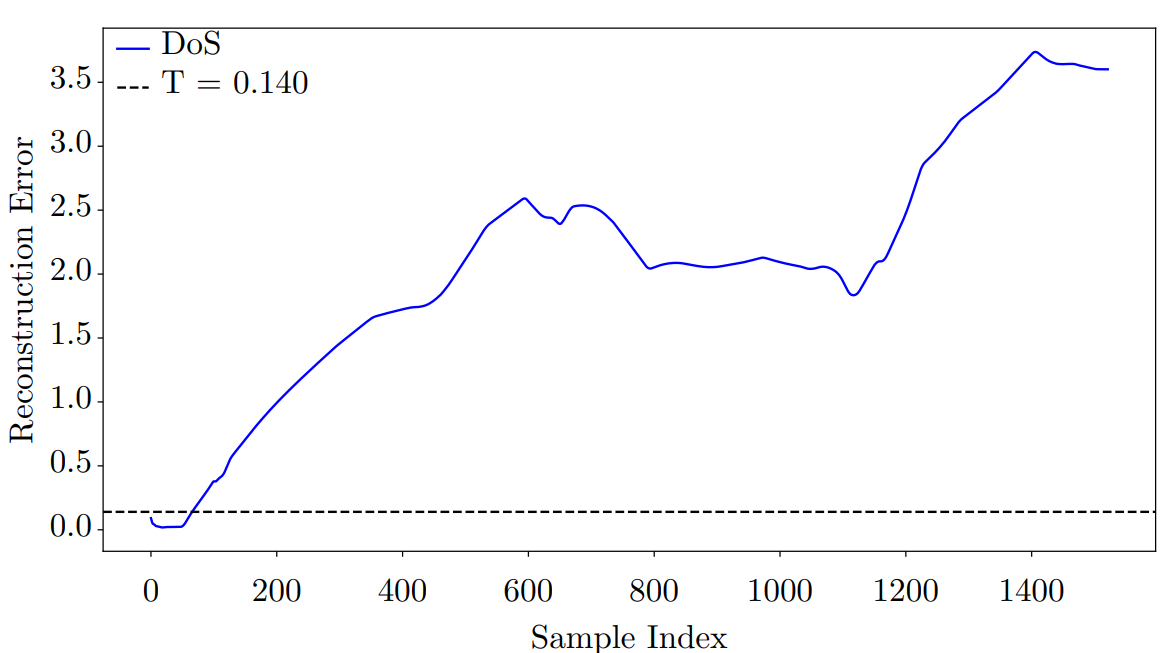}
    \caption{Detection performance for DoS.}
    \centering
    \label{fig:dos2}
\end{figure}

To imitate a denial-of-service attack on the controller-battery signal, we force the signal to always be zero. This results in the battery never charging or discharging in a 24 hours period and remaining in a constant state. However, this behavior is in violation of the expected normal and operational behavior of a EC. Figure~\ref{fig:dos1} shows the distribution of reconstruction error for this variation of attack while Figure~\ref{fig:dos2} plots the detection performance of the model. The results show that the model is good at detecting this attack, with a high detection rate and only a few false negatives.

\subsubsection{Power Injection Attacks}


In PI attacks, an adversary aims to discharge the battery in an unsafe and unexpected manner to inject maximum possible power. A battery normally discharges when it supplies power to connected loads. However, an adversary can also artificially force the battery to discharge, essentially making the battery act like it is supplying power to loads. This discharging action results in the battery injecting its stored power. To simulate PI attacks, we apply negative scaling of different proportions of the original value to the controller-battery signal. 

\begin{figure}[htbp!] 
    \centering 
    \includegraphics[width=1\textwidth]{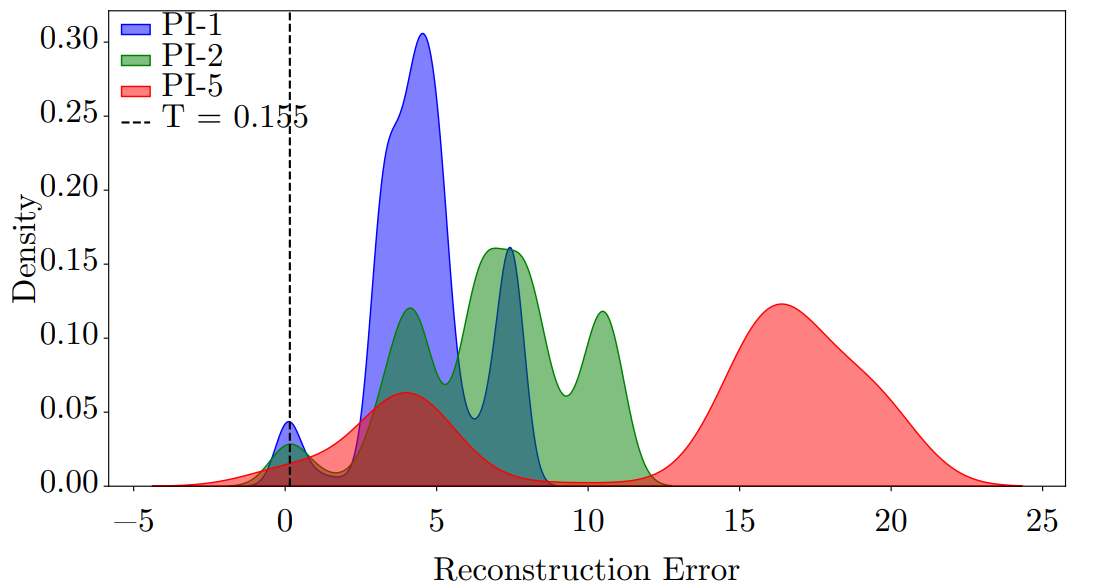}
    \caption{Distribution of Reconstruction error for PI attacks.}
    \centering
    \label{fig:PI1}
\end{figure}

\begin{figure}[htbp!] 
    \centering 
    \includegraphics[width=1\textwidth]{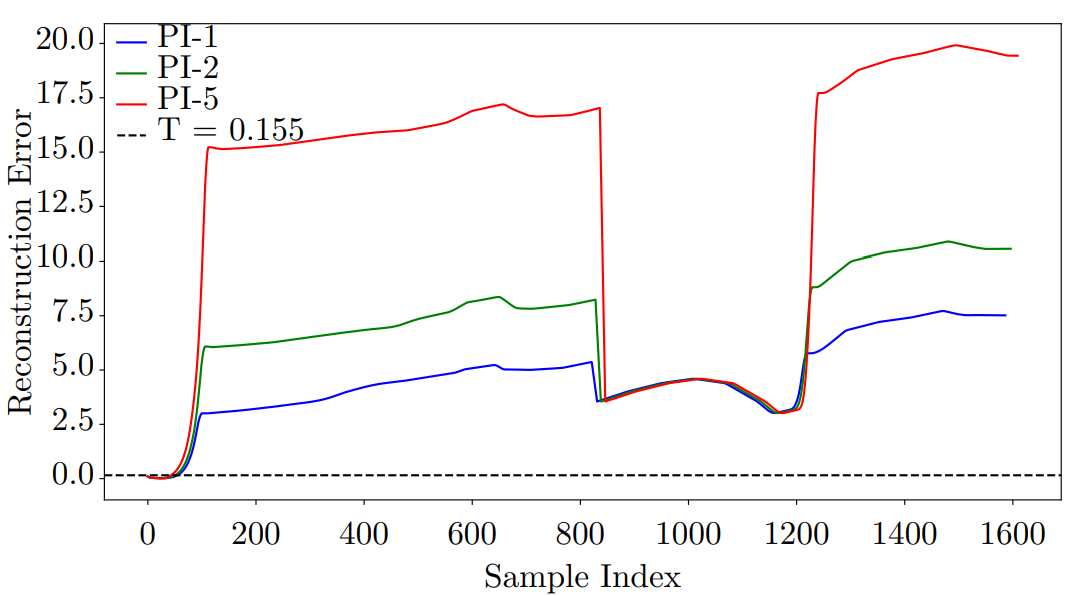}
    \caption{Detection performance for PI attacks.}
    \centering
    \label{fig:PI2}
\end{figure}

For these attacks, we experiment with three different variations. The first attack simply reverses the input signal. The second reverses and amplifies the signal by twice while the third reverses and amplifies the signal by five times. Figure~\ref{fig:PI1} shows the distribution of reconstruction error for PI attack samples. The dotted line (T) indicates the detection threshold. Figure~\ref{fig:PI2} depicts the detection performance of the model. In the figures, PI-1 represents reversal of the original value with same magnitude, while PI-2 and PI-5 represent reversal and amplification by a factor of two and five, respectively. The model shows good performance for all variations of this attack with the observation that the higher the deviation from normality (higher the amplification), the higher the reconstruction error, and the easier it is for the model to detect the attack. 

\subsubsection{Load Reduction Attacks}


Load altering attacks imitate the scenario where an attack increases or decreases demand of the controllable loads within the households. LR attacks result in a reduction of demand. We simulate these attacks using signal disturbance in Simulink so that within the specified time range defined by start and duration parameters, output is set to a factor of the original value. Outside the specified time range output is kept at 1.0. 

Two variations of LR attacks are studied. First, we scale down the input signal to half of the original value (LR x0.5), resulting in significant reduction in power consumption. To imitate an attacker taking over multiple flexible assets, both the households have their demand reduced at the same time. Secondly, we force the demand to be 0 (LR x0) resulting in unexpected behavior as in reality the demand is never exactly 0.

\begin{figure}[htbp!] 
    \centering 
    \includegraphics[width=1.0\textwidth]{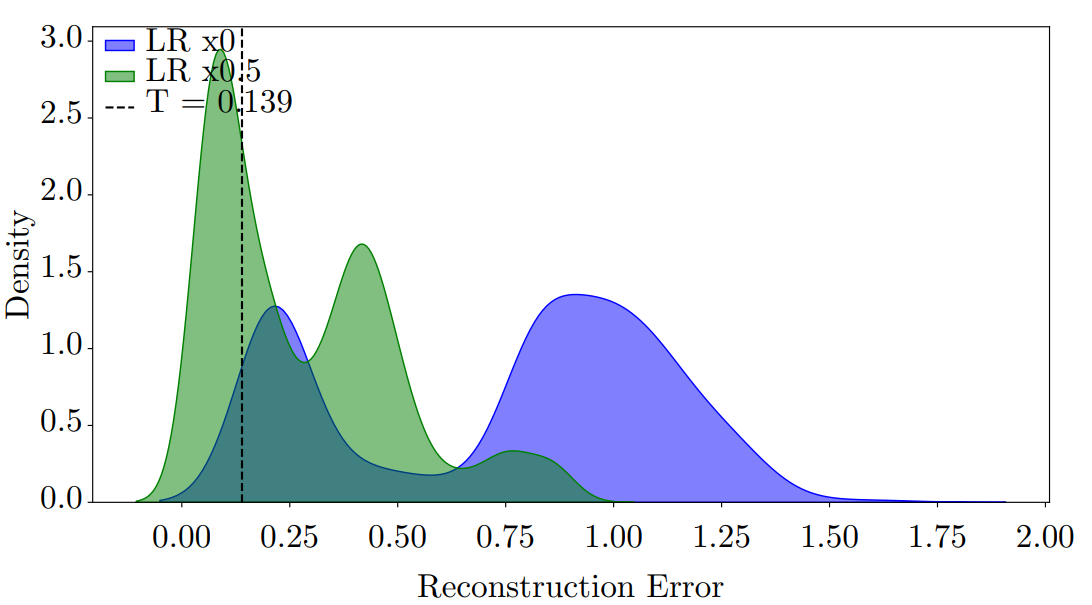}
    \caption{Distribution of Reconstruction error for LR attacks.}
    \centering
    \label{fig:LR1}
\end{figure}

\begin{figure}[htbp!] 
    \centering 
    \includegraphics[width=1.1\textwidth]{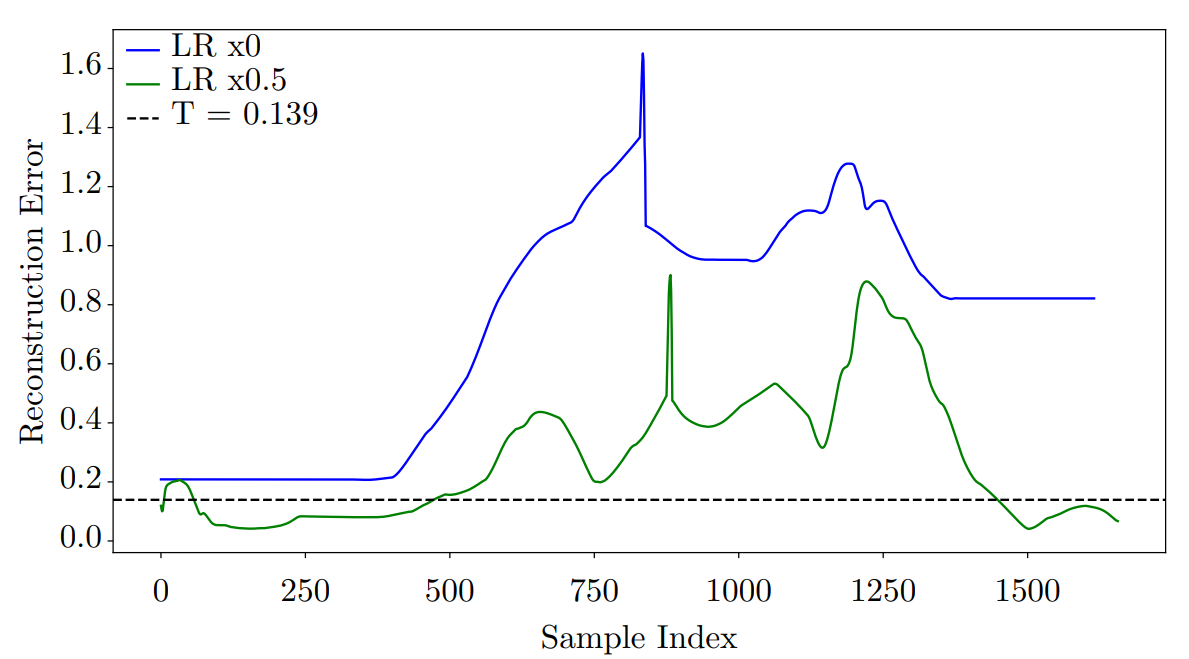}
    \caption{Detection performance for LR attacks.}
    \centering
    \label{fig:LR2}
\end{figure}

Figure~\ref{fig:LR1} shows the distribution of reconstruction error for LR attack samples. The dotted line (T) indicates the detection threshold. Figure~\ref{fig:LR2} depicts the detection performance of the model. The model shows good detection performance for both variations. 

\subsubsection{Load Increase Attacks}

LI attacks are represented by abruptly amplifying the consumption or demand of controllable loads within households from the normal value. This results in the grid needing to supply unexpectedly extra power to meet the increased demand. Two variations are investigated for these attacks where first the demand is doubled (LI x2), and then the demand is increased by a factor of five (LI x5).

\begin{figure}[htbp!] 
    \centering 
    \includegraphics[width=1.0\textwidth]{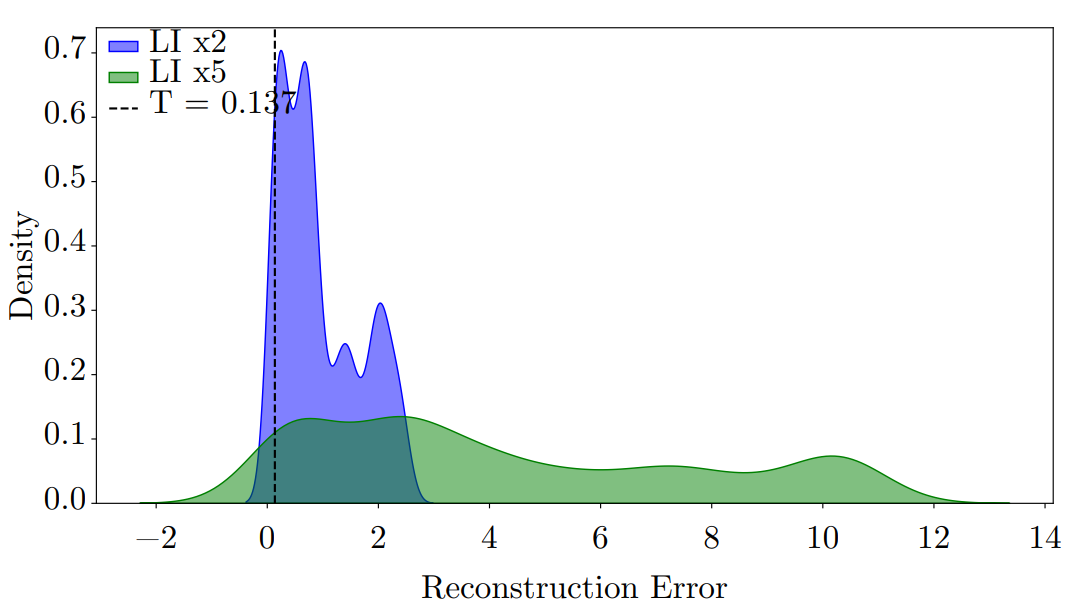}
    \caption{Distribution of Reconstruction error for LI attacks.}
    \centering
    \label{fig:LI1}
\end{figure}

\begin{figure}[htbp!] 
    \centering 
    \includegraphics[width=1.1\textwidth]{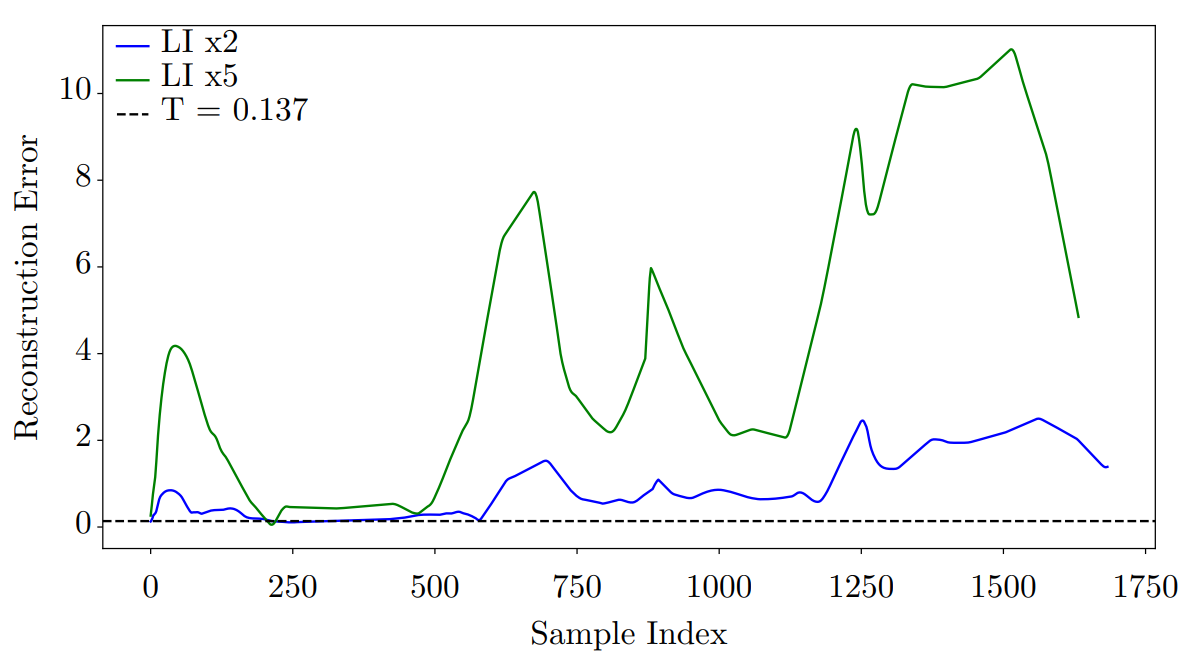}
    \caption{Detection performance for LI attacks.}
    \centering
    \label{fig:LI2}
\end{figure}

Figure~\ref{fig:LI1} shows the distribution of reconstruction error for LI attack samples. The dotted line (T) indicates the detection threshold. Figure~\ref{fig:LI2} depicts the detection performance of the model. The model shows good detection performance for both variations. 

\subsubsection{Other Metrics}

To evaluate the detection performance of the model using metrics such as precision, recall, and F1-score, we collect an additional test dataset. This dataset includes both normal and abnormal data, enabling calculation of these metrics. Table~\ref{tab:conclusion_evaluation} summarizes the evaluation metrics for different attacks. 











\begin{table}[htbp!] 
\centering
\caption{Summary of evaluation results.}
\begin{tabular}{| p{0.2\linewidth} | p{0.2\linewidth} | p{0.2\linewidth} | p{0.2\linewidth} |}
\hline
\textbf{Attack Type} & \textbf{Precision} & \textbf{Recall} & \textbf{F1-score}
\\ \hline
PA x2 & 0.6490 & 0.1317 & 0.2189 
\\ \hline
PA x5 &  0.8505 & 0.4086 & 0.5520 
\\ \hline
DoS &  0.9240 & 0.9507 & 0.9372 
\\ \hline
PI-1 &  0.9280 & 0.9660 & 0.9466 
\\ \hline
PI-2 &  0.9286 & 0.9693 & 0.9485 
\\ \hline
PI-5 & 0.9295 & 0.9739 & 0.9512 
\\ \hline
LR x0 &  0.9103 & 0.7469 & 0.8205 
\\ \hline
LR x0.5 &  0.8883 & 0.5709 & 0.6951 
\\ \hline
LI x2 &  0.9249 & 0.8705 & 0.8969 
\\ \hline
LI x5 &  0.9314 & 0.9896 & 0.9596 
\\ \hline
\end{tabular}
\centering
\label{tab:conclusion_evaluation}
\end{table}



\subsection{Collaborative Learning}

The centralized autoencoder model as discussed thus far performs well but depends on raw data to train on and learn the normal operational patterns. Since the data can be sensitive and belongs to private individuals, this dependency can hinder its real world potential. To assess the approach in a more practical and privacy-preserving setting, we train an autoencoder using FL. We use the Flower framework~\cite{flower} to implement the FL process. We experiment with a configuration of three clients that communicate with a server using the Flower API. The server is configured to run the training process for three full rounds where training in each round takes the model closer to convergence and losses reduce. The server aggregates model updates from clients using the FedAvg~\cite{fedavg} algorithm. 

\begin{figure}[htbp!] 
    \centering 
    \includegraphics[width=1.0\textwidth]{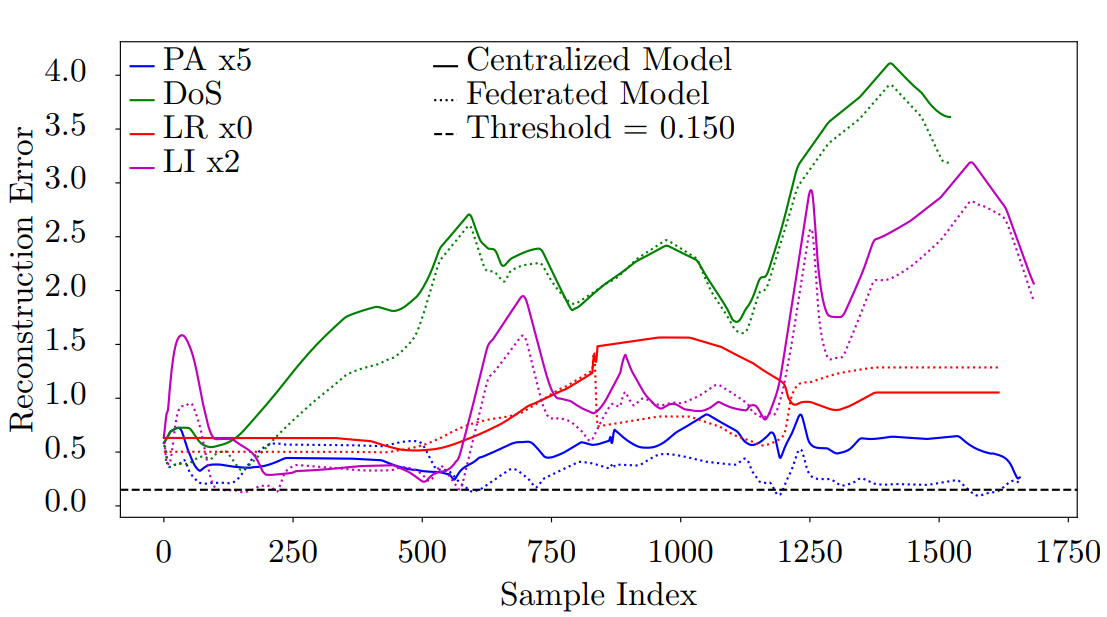}
    \caption{Comparison of centralized and federated model performance.}
    \centering
    \label{fig:FL1}
\end{figure}

Once the FL process is complete, we use the final global model to detect the same attacks as the centralized autoencoder model in Section~\ref{results}. Figure~\ref{fig:FL1} shows a comparison between the performance of the centralized and federated models for a selection of attacks. The federated model shows comparable performance to the centralized model for all attacks. This motivates the potential for using the proposed approach in a real-world setting, as it ensures data privacy.

\section{Interpretation of Results}

The aim of this paper is to enhance the security of ECs by proposing and demonstrating the effectiveness of an IDS in such environments. A review of the literature revealed that there is a lack of thorough research in the area of security for ECs, especially in detection methods. To address this gap, we propose a detection system based on deep autoencoders, which is a type of unsupervised model. The results of the study demonstrate the potential of the proposed IDS for ECs. The detection algorithm is capable of identifying both economically damaging attacks and those that may lead to unsafe working conditions, highlighting its relevance for real-world deployment. These findings confirm that anomaly-based intrusion detection can be a valuable tool in enhancing the cybersecurity and resilience of energy infrastructures.

The selection of autoencoders as the detection model among other possibilities is mainly motivated by previous research which has found autoencoders to be effective in many cyber-physical settings. For the architecture of autoencoders, there are many possibilities and optimizations that can be applied. Our goal was to achieve a fine balance between model complexity and performance and that is the reason we implemented a model of five layers (two encoding, one repeat, and two decoding). The selection of the optimizer, loss function, number of epochs, detection threshold, and other hyperparameters are also based on the goal of keeping the model simple. While, we have demonstrated that the performance of the model is good for most attack scenarios, these hyperparameters can be fine-tuned to further enhance the performance. 

As data-driven detection systems are as good as the data they train on, we made an attempt to build a model in Simulink that is accurate, well-tuned, and based on real-world patterns (e.g., consumption trends). While this simulated data can not fully replace a real world data, it can be considered very close to it. In a field where data from real systems is impractical to access for many legal and data protections reasons, simulated data enable us to design and test our detection systems. However, real-world testing of the system is necessary for final validation and deployment. 

We have also considered privacy concerns that such a detection system can suffer from in a real-world  setting. The detection system relies on raw measurements about voltages, currents, and power from private households, and that can prove to be a big challenge in reality. To address this concern, we have demonstrated that the detection model can be trained in a collaborate manner using FL. While the decentralized approach slightly reduces detection accuracy, it ensures that sensitive data remains local, mitigating privacy risks. This may introduce a trade-off between performance and privacy but finding the right balance is particularly important in ECs, where data privacy concerns can hinder the adoption of centralized security solutions. Despite the minor decrease in performance, the advantages offered by FL, such as improved data confidentiality and compliance with privacy regulations, make it a viable and valuable approach. 

Nonetheless, there is room for further improvement. One of the main challenges observed in the results is the presence of false negatives, specially for PA attacks. Future research should focus on refining the detection algorithm to enhance performance, particularly by improving model recall without reducing the precision. This could be achieved through more sophisticated feature selection, more complex deep learning models, or integrating more operational data, e.g., network data, into the training phase. The performance of the federated model could also be further improved by finding optimal hyperparameters such as number of rounds and number of epochs per training round. However, this needs to be carefully observed as increasing the number of epochs allows each client to train more on its local data in each round which may improve the performance, but it may lead to overfitting if the data are small. Similarly, increasing the number of rounds allows the global model to improve over time, but it comes at the cost of increased communication overhead. Overall, the results indicate that the proposed approach is a promising step toward secure and privacy-preserving anomaly detection in ECs. Continued research and optimization efforts will be essential to ensure its robustness and practical applicability in dynamic and complex environments such as ECs.




\section{Conclusions}

This paper presents an anomaly-based intrusion detection system developed to enhance the security of energy communities while preserving data privacy. By leveraging deep autoencoders and unsupervised learning, the proposed model effectively detects cyber threats that could lead to economic losses or hazardous conditions. The results demonstrate good detection performance, confirming the potential of the approach for real-world application. The incorporation of federated learning, while introducing a slight reduction in performance, successfully ensures data privacy and decentralization. This approach provides a strong foundation for future development, particularly in scenarios where privacy is a key requirement. Our work represents a promising advancement in cybersecurity for energy communities. With further refinements and real-world validations, the proposed detection system has the potential to become a vital component of secure decentralized energy infrastructures. 


\backmatter

\section*{Declarations}

\subsection*{Competing interests}
The authors declare that they have no competing interests.

\subsection*{Funding}

The work was conducted as part of the Cybersecurity for Resilient Energy Communities of the Future (CyREC) project, funded by the Swedish innovation agency (reference number 2023-02987), and the RAISE project under the MUR National Recovery and Resilience Plan funded by the European Union - NextGenerationEU. The funderx had no role in the design of the study; in the collection, analyses, or interpretation of data; in the writing of the manuscript; or in the decision to publish the results.


\subsection*{Author contribution}

Writing-original draft, Z.A, G.G; conceptualization, Z.A, G.G, M.A; methodology, Z.A, G.G; funding acquisition, M.A. All authors have read and agreed to the published version of the manuscript.

\subsection*{Data availability}
The code and data used in this work are openly available in the repository at~\cite{repo}.


\subsection*{Ethical approval}
Not applicable.

\bibliography{references.bib}

\end{document}